\documentstyle{article}
\begin{document}
\title{\Large \bf  Local Magnetic Moments in Nanoscale
         Systems\\}
\author{\large Liang-Jian Zou \\
       {\it    Institute of Solid State Physics, Academia Sinica,
               P.O.Box 1129, Hefei 230031, China  } \\
               X. G. Gong and Qing-Qi Zheng       \\
       {\it    CCAST (World Laboratary) P.O.Box 8730, Beijing, 100080 and
               Institute of Solid State} \\ {\it Physics, Academia Sinica,
               P.O.Box 1129, Hefei 230031, China  } \\
               C. Y. Pan  \\
       {\it    Center of Magnetism in Information technology, Department of
               Physics, Utah State} \\{\it University, Logan, UT 84322-4415 } \\ }
\date{ }
\maketitle

\large
\begin{center}
                            ABSTRACT \\
\end{center}
  The formation of local magnetic moments and its size effect
in one- and three-dimension finite
systems with magnetic impurity are investigated based on the Anderson
hybridizing model in real space. By the exact diagonalization within the
self-consistent mean field
approximation, it is found that as contrast to that in infinite system, the
formation of the local magnetic moment in finite system
may strongly depend on the size of system up to a few
nanometers. The spectral densities of the local electronic states and the
spatial distribution of the magnetic moments for the conduction electrons
polarized by
the local moment are also obtained. The condition for the occurrence of local
magnetic moments in such small systems is
discussed. The strong size-dependence of the local magnetic
moments in finite system is attributed to the quantum size effect.

\vspace{7 mm}


\newpage
\begin{center}
{\bf I. INTRODUCTION}
\end{center}

   The magnetism of small particles has attracted great interests both in
the theoretical and in the experimental study [ 1 - 7 ].
Nanosized magnetic particles are fundamentally important in
cluster and nano-crystal physics and have extensive potential application.
Experiments [ 1 - 3 ] have demonstrated that the magnetic
properties of the 3d or 4d transition metal and the rare-earth metal clusters
are different from those of the bulk materials. For example, the magnetic
moments decrease with respect to the increase of cluster sizes,
the ferromagnet-paramagnet transition temperature
(Curie temperature) of the 3d transition metal clusters shows strong
size-dependence [ 1 - 3 ]. Some 4d transition metal clusters may exhibit
giant magnetic moments [ 4 ], though their bulk materials are nonmagnetic.

    The theoretical study of the size effect on the magnetic properties of the
transition metal and the rare-earth metal clusters has become an interesting
subject [ 5 - 9 ]. Pastor et al. [ 5 ] studied
the influence of the electron-electron correlation and the cluster geometry on
the magnetism in the framework of the tight-binding Hubbard model. They found
that the local magnetic moments of pure 3d transition-metal clusters,
Fe$_{n}$, Cr$_{n}$ and Ni$_{n}$ (n $\ <$ 10), have weak size-dependence,
their prediction for the size-dependence of the magnetic moments
of Fe$_{n}$ clusters disagrees with the experimental results [ 1 ].
Recent electronic structure calculation performed by Reddy et al.
[ 6 ] for pure 4d transition-metal clusters, Pd$_{13}$, Ru$_{13}$ and
Rh$_{13}$, shows
the existence of the magnetic moment and magnetic order in Rh$_{13}$
cluster, but the predicted magnetic moment per atom is larger than those found
in the experiment [ 4 ]. This discrepancy might arise from the unreal
geometries for the clusters used in the calculation. Jena et al. [ 7 ] studied
the Ising model for small magnetic clusters at finite temperature by using
the Monte-Carlo technique. Linderoth et al. [ 8 ] discussed the finite size
effect on the spin-wave spectrum
in magnetic clusters in terms of the Heisenberg exchange model with structural
relaxation, they showed that the spin-wave spectrum in these finite clusters is
very different from that in the bulk cases, though the definition
of the spin wave in such small systems is quite ambiguous.
From these studies, it is clear that the magnetic properties in the finite
systems are different from those in bulk phases.

    The main subject of the studies mentioned above concentrates on the
average magnetic moment, the
magnetic order and the exchange interaction in magnetic clusters,
the existence of local magnetic moment and its stability in nano-system
are not deeply explored. Although some work on local moment of magnetic
impurity in nonmagnetic clusters are investigated by several authors [ 10,
11 ], the
general behavior of the formation of local magnetic moment in finite system is
not clear. In the first-principle electronic structure calculation for
FeR$_{n}$ clusters (R=Al, Mg, Ca, Sr, etc.), Ellis and Guenzburger
[ 10 ] found that Fe impurity is nonmagnetic for FeAl$_{42}$, while
it is magnetic for FeMg$_{26}$, FeCa$_{42}$, and FeSr$_{18}$.
Using the local spin density method, Callaway et al. [ 11 ]
studied the electronic structure of Mn impurity in fcc Al matrix (a free
MnAl$_{18}$ cluster), the Mn atom in the cluster is found to be magnetic, the
magnetic moment is about 1.78
$\mu_{B}$, which is much smaller than its atomic moment. But how and when the
local magnetic moments can form in such small systems are not clear yet.

   The picture for the formation of local magnetic moment in bulk metal (
infinite system) had been established by Friedel, Anderson and many other
authors [ 12 - 17 ] in the early of 1960s, they showed that an impurity
dissolved in bulk
metal may form Friedel's resonant states [ 12, 13 ] (or virtual bound-states)
with the conduction electrons through
the hybridizing interaction, the spin-split of the Friedel's resonant states
leads to the formation of local magnetic moment. It is well known [ 16, 17 ]
that the condition for the occurrence of local magnetic moment
in bulk is U$\rho_{d\sigma}(E_{F})$ $\ > $ 1, where $\rho_{d\sigma}(E_{F})$
is the densities of states for the d-electrons at the Fermi surface
and U is the on-site Coulomb interaction of the local d-electrons, the energy
level of the d-electrons of the impurity with respect to the Fermi
energy, $\epsilon_{d}$, must be negative so as to form the local
moment, and it is size-independent. However,
when the size of bulk metal is reduced to just a few
nanometers, does this condition still hold? Obviously, in this situation
the quantum size effect will play an important role in
the formation of local magnetic moment.

   In this paper, we have studied the finite size effect on the formation of
local magnetic moment by using the Anderson impurity model. It could provide
the understanding on the magnetic behavior of small systems and the evolution
of the magnetic properties from atom, through small cluster to bulk system.
The rest of this paper is arranged as following: the formalism of the local
magnetic moments in finite systems
is described in Sec.II, the results and discussions for one-dimension (1D)
and three-dimension (3D) cases are given in Sec. III, and
the conclusion is drawn in Sec.IV.

\begin{center}
{\bf II. FORMALISM}
\end{center}

  A magnetic impurity, such as a transition-metal atom
dissolved in a nonmagnetic metal, can be described by the Anderson
hybridization Hamiltonian [ 12 ]. The Anderson model
provides an essential description for the mechanism of the formation of
a magnetic moment in solid, and plays an important role in our understanding
on the existence of local magnetic moment in infinite system. This model is
believed to be still valid in describing the physical process in the finite
systems. 
In real space, the Hamiltonian of the Anderson hybridizing
model with a magnetic impurity at site R$_{i}$ can be expressed in matrix form:
\begin{equation}
  H = \sum_{\sigma}[C^{\dag}_{\sigma} {\bf T} C_{\sigma} + (\epsilon_{d}+
      \frac{U}{2} \sum_{\sigma '}d^{\dag}_{\sigma '}d_{\sigma
'})d^{\dag}_{\sigma}d_{\sigma}
      +C^{\dag}_{\sigma} {\bf P}d_{\sigma}+ d^{\dag}_{\sigma} {\bf P^{\dag}}
      C_{\sigma}]
\end{equation}
where C$^{\dag}_{\sigma}$=
(c$^{\dag}_{1\sigma}~\dots~c^{\dag}_{i-1\sigma}~0~$
c$^{\dag}_{i+1\sigma}~\dots~c^{\dag}_{N\sigma})$ denotes the creation
operator matrix of the mobile (or conduction) electrons
with spin ${\it \sigma}$, d$^{\dag}_{\sigma}$ the creation operator of the
{\it d}-electron of the impurity, and $N$ is the electron number
in the system.  ${\bf T}$ represents the hopping matrix of the conduction
electrons, which is a {\it N} $\times$ {\it N} matrix, it has
simple tridiagonal form for 1D chain:
\begin{equation}
    {\bf T} = \left( \begin{array}{ccccc}
\epsilon_{c} & t & 0 & \cdots & 0 \\
t & \epsilon_{c} & t & \cdots & 0 \\
\vdots & \vdots & \vdots & \ddots & \vdots  \\
0 & 0 & 0 & \cdots &  \epsilon_{c}  \end{array} \right)
\end{equation}
However it is more complicated in the case of higher dimensional systems, and
it strongly depends on the structure of the small clusters. In Eq.(2), the
matrix element $t$ represents the hopping energy of the conduction electron
from one site to its nearest-neighbor
site; $\epsilon_{d}$ (or $\epsilon_{c}$) denotes the difference between the
impurity (or the conduction) electron energy level $E_{d}$ (or $E_{c}$) and
the chemical
potential $\mu$, and U is the on-site Coulomb interaction.
${\bf P}=(~\dots~v_{i-1}~0~v_{i+1}~\dots)$ refers the hybridizing matrix
between the d-electron and the conduction electrons. In the present paper, only
the nearest-neighbor interaction is considered. In finite system,
the movement of the conduction electron can't be considered as
a plane wave, therefore we can not describe the impurity effect by the usual
scattering theory and the phase shift technique. In this paper, the real-space
Green's function
technique is employed to deal with the impurity effect in finite system.
Within the self-consistent mean field approximation, by introducing the mean
occupancy of the d-electrons, $ \bar{n}_{\sigma}$ =
$<d^{\dag}_{\sigma}d_{\sigma}> $,
the propagator of the d-electron and the mixed propagator matrix between the
conduction electrons and the d-electron are given as the follows:
\begin{equation}
  <<d_{\sigma};d^{\dag}_{\sigma}>> =
[\omega -({\epsilon}_{d} + U \bar{n}_{\bar{\sigma}}) - {\bf P} (\omega {\bf I}
-{\bf T})^{-1} {\bf P}^{\dag}]^{-1}
\end{equation}
\begin{equation}
  <<C_{\sigma};d^{\dag}_{\sigma}>> = [\omega
{\bf I} -{\bf T}]^{-1}  {\bf P}^{\dag}  <<d_{\sigma};d^{\dag}_{\sigma}>>
\end{equation}
where ${\bf I}$ is the unit matrix. The propagator matrix of the conduction
electrons is:
\begin{equation}
  <<C_{\sigma};C^{\dag}_{\sigma}>> = [\omega {\bf I} -{\bf T}-
{\bf P} \times  {\bf P}^{\dag}/(w-\epsilon_{d}-U n_{-\sigma}) ]^{-1}
\end{equation}
The occupation number $n_{d}$ and the magnetic moment $m_{d}$ of the
d-electrons can be obtained through the preceding propagators:
\begin{equation}
    n_{d} =\frac{1}{\pi} \int d \omega f(\omega) Im [
{<<d_{\uparrow};d^{\dag}_{\uparrow}>>}_{\omega-i\eta} +
{<<d_{\downarrow};d^{\dag}_{\downarrow}>>}_{\omega-i\eta} ]
\end{equation}
and
\begin{equation}
    m_{d} =\frac{1}{\pi}  \int d \omega f(\omega) Im [
{<<d_{\uparrow};d^{\dag}_{\uparrow}>>}_{\omega-i\eta} -
{<<d_{\downarrow};d^{\dag}_{\downarrow}>>}_{\omega-i\eta} ]
\end{equation}
Where f$(\omega)$=1/(exp($\beta$ ($\omega$-$\mu$))+1) is the Fermi-Dirac
distribution function. The total spectral
density of the excited states for the local electrons is:
\begin{equation}
    \rho_{d}(\omega) = \frac{1}{\pi} Im [
{<<d_{\uparrow};d^{\dag}_{\uparrow}>>}_{\omega-i\eta} +
{<<d_{\downarrow};d^{\dag}_{\downarrow}>>}_{\omega-i\eta} ]
\end{equation}
Once upon a local magnetic moment forms, its surrounding conduction electrons
will be polarized, the magnetic moment of the polarized conduction
electrons at site R$_{i}$ can be expressed as:
\begin{equation}
    m_{s}(i) =\frac{1}{\pi}  \int d \omega f(\omega) Im [
{<<c_{i\uparrow};c^{\dag}_{i\uparrow}>>}_{\omega-i\eta} -
{<<c_{i\downarrow};c^{\dag}_{i\downarrow}>>}_{\omega-i\eta} ]
\end{equation}
The occupation number and the magnetic moments are obtained by the
self-consistent mean field approximation. Although the mean field approach is
somewhat simple and approximate, it might produce
some misleading results at finite temperature, however at T=0 K, it indeed
provides
many physical results [12,17]. For example,  the mean-field results for the
formation of local magnetic moments give a good qualitative description of an
impurity in a nonmagnetic metal, so it remains sense near the temperature of
absolute zero degree. As the first step, the mean-field approximation is a
good approach
to outline tthe basic picture of the local magnetic moment in finite systems.

   In infinite system, the matrix product
${\bf P}(\omega {\bf I} - {\bf T})^{-1} {\bf P}^{\dag}$ in Eq.(3) can be
expressed in terms of a summation over the wavevectors of the conduction
electrons. However, in finite case, the translation symmetry is broken,
the evolution of the local magnetic moment with respect to the size of
systems has to been studied numerically. In this paper, the closed set
of the self-consistent equations (3) to (6) are solved by exact
diagonalization to calculate the Green's function and the local spectral
density. In the following, all the energies
are measured in units of the hopping integral $t$ and the magnetic moment
in units of the Bohr magneton, $\mu_{B}$.

\begin{center}
{\bf III. RESULTS AND DISCUSSIONS}
\end{center}

The formation of a local magnetic moment and its dependence on the size of the
finite system are studied in this section.
To explore the explicit interrelation of the local magnetic moment, the
cluster size and the dimensionality of the system, we first consider
a linear quantum chain with an impurity situated at the center at
T=0 K, the formation of a local magnetic moment in 3D system with an impurity
will be discussed later. The maximum numbers
of the atoms in the systems we have studied are 30 for 1D case, and 15 for
3D case, respectively, which correspond to several nanometers.

  From the numerical calculations for various systems with different parameters,
we find two points: (a) When the impurity energy level $\epsilon_{d}$ is far
below the chemical potential and the Coulomb interaction U is strong enough,
the local moment can form and is almost not
affected by the variation of cluster size. (b) If the energy level
$\epsilon_{d}$ approaches or lies above the chemical potential $\mu$, and the
Coulomb interaction is not strong enough, there will be no local moment. In
this situation,
the impurity level is too shallow to form localized Friedel's resonant
states, or the Coulomb interaction is too weak to lead to the spin-split of
the Friedel's resonant states.
These two points are similar to the results of infinite systems [ 12 ].
These results are reasonable, since the former corresponds to the isolated atom
limit when the impurity level is deep and the on-site
Coulomb interaction is strong enough, the magnetic moment of the impurity is
almost not affected by the environment; and the latter corresponds to the free
electron limit. These results are valid both for 1D quantum chains and for 3D
nano-clusters. However, there are also other distinct and interesting behavior
when the finite systems lie between these two extreme cases, we will address
it in detail in the following.

\vspace{0.5cm}

\noindent{\bf III.1 One-Dimensional Finite Systems}

\vspace{0.5cm}

  To avoid the effect of the asymmetry on the formation of the local magnetic
moment in the cases of even-number-particle systems, we consider only the
systems with odd-number atoms.
Fig.1 and Fig.2 show the local magnetic moment as a function of the size
for 1D linear quantum chains with different parameters. A strong dependence of
the
magnetic moment on the number of the atoms can be seen clearly. It is found
that the value
of the magnetic moment mainly depends on three factors: the system size, the
impurity energy level and the on-site Coulomb interaction. The
local moment will approach a constant (the bulk value) when the cluster size
increases. A non-monotonous descent of the local magnetic moment with respect
to the increase of the size can be seen in Fig.2
for the 1D systems with small Coulomb interaction and shallow energy level.
The shallower the impurity energy level is, the stronger the variation of the
local magnetic moment is. The d-electron will be delocalized and lose
its magnetic moment when the energy level becomes shallow and the size of the
system becomes large enough (See the curves (b) and (c) in Fig.1).
The reason is that the impurity electron would be influenced by
the surrounding electronic field, when more
nonmagnetic metal atoms, correspondingly more conduction electrons,
are added to the finite system, the local environment of the system is changed,
the modification to the electrons in shallow energy level is stronger than that
in deep energy level. In the mean time, because of the
quantum size effect, the difference between the
energy levels and the density of states for the conduction electrons are also
changed, therefore for the small system, the Friedel's
resonant states, hence the formation of the local magnetic moments is very
sensitive to the local environment.

   So far there is no any direct experimental data on the magnetic properties
of the "dilute clusters" available for comparison with the present results.
We have noted that similar behavior of the local electronic structure
is observed experimentally for Co impurity in nonmagnetic
matrix, such as in CoAl$_{n}$ (n = 1-35) [ 18 ] clusters. It reflects that the
local environment of the magnetic impurity changes with the variation
of the nonmagnetic atoms in cluster.
As seen from Fig.2., the delocalization of the d-electron happens abruptly
rather than in a gradual way [ 1, 2 ]. This can be attributed to the absence of
the Coulomb correlation between d-electrons at different sites and
the less coordinate number in 1D case.

  The spectral density of the excited states for the d-electrons can be
obtained
from the propagators in Eq.(8). The total spectral densities of the excited
states in the systems with N=15 and N=25 are shown in
Fig.3 and Fig.4, respectively. There exist
two local modes near the lower energy level $\epsilon_{d}$ and upper energy level
$\epsilon_{d}+U$, these two local modes correspond to the single-electron levels.
Similar to the infinite systems, there are a series of closely
spaced states lying between them, which may form a narrow hybridizing band
when N approaches infinity, these two bands might correspond to the Friedel's
resonant states, the fact that the levels are discrete and the width of the
resonant states is very narrow implies that the finite system exhibits
strongly localized character. The quantum size effect of the energy level
cann't seen clearly from the Fig.3 and Fig.4 since the energy-level shift is
relatively small, it is about 0.001 to 0.01, which is difficult to reflect
in the figures. It is found that the electrons most probably
occupy the spin-up lower energy level, and the present distribution
of one-electron energy levels has some similarity to
that obtained by the local density functional technique [ 9 ].

   When a local magnetic moment is situated at a nonmagnetic metal, it
will polarize the
spins of its surrounding conduction electrons, therefore the local moment will
be partially or completely "screened" by the conduction electrons. An
exactly screened local moment by the conduction electrons will form a singlet
and may lead to the abnormality of the conductivity and the specific heat. This
phenomenon is the
so-called "Kondo effect" [ 19, 20 ]. Recently, the size-dependence of the
Kondo energy scale, i.e. the Kondo temperature, is studied both in
point contacts [21] and in thin film, an explicit size-dependence of the
Kondo temperature and the Kondo peak in the differential resistance is observed
in [21]. The magnetic moment distribution of the polarized conduction electrons
around the local magnetic
moment is an important physical properties in such a dilute magnetic cluster.
We have calculated the spatial distribution of the spins of the conduction
electrons for two systems with N=15 and N=25.
The distribution of the polarized magnetic moments is shown in Fig.5. The
nearest-neighbor polarized spins are antiparallel to the local moment, it may
be parallel in the next-nearest-neighbor site. the nearest-neighbor
polarized moment
in the 1D system is -0.09 $\mu_{B}$ for N=15 and -0.12 $\mu_{B}$ for N=25.
It is found that the period and amplitude of the Friedel's oscillation change
with respect to the size, although these two systems ( with different sizes
N=25 and N=15) has the same parameters. It is found that the larger the system
is, the shorter the period of the Friedel's oscillation is. Because of the
hybridization interaction between the conduction electrons and the d-electron,
the conduction electrons have some possibilities to enter the impurity atom, so the
number and the magnetic moment of the conduction electrons don't vanish at the
impurity site.

\vspace{0.5cm}

\noindent{\bf III.2 Three-Dimensional Finite Systems}

\vspace{0.5cm}

   In the following, we have studied the formation of the local magnetic
moment and the size effect on
the magnetic properties in 3D dilute clusters,
however, for the 3D cases, one may meet the difficulty of the giant possible
configurations when the number of atoms in clusters exceeds 8-10 [ 5, 9 ].
The real geometry must be determined by other methods, such as the
first-principle electronic structure calculation by minimizing
the total energy or the molecular-dynamic simulation by ensuring
the structure in the lowest equilibrium state. To determine the optimal cluster
structure is beyond our scope. In
the present paper, we investigate some small 3D systems with the high-symmetry
structure, these systems consist of 5, 7, 9 and 15 atoms. The corresponding
structures of these systems
are chosen as tetrahedron for 5-atom clusters, octahedron for 7-atom clusters,
cube for 9-atom clusters and 6-capped-cube for 15-atom clusters. The impurity
is situated at the center of the clusters, and the nonmagnetic metal atoms
surround it.

   A typical size-dependence of the local magnetic moments in 3D finite systems
is shown in Fig.6. The local magnetic moments decrease monotonously with the
increasing atoms when $\epsilon_{d}$ is shallow and U is small.
Comparing with the 1D case, the local magnetic moment
in 3D systems varying smoothly with the size may result from the increase of
the coordination number. Though the absence of the valley feature in
3D cases, the
dependence of the local magnetic moment on the cluster size are obvious. The
strong size-dependence of the formation of the local magnetic
moments originates from the quantum size effect. Because in these situations,
the mean coordination number and the overlap of the wavefunction of
atoms are reduced, which lead to the narrowing of the d-electron and the
conduction electron bands, the localized character become
important for these finite systems. From Fig.6, one also finds that the
impurity atom becomes nonmagnetic if the
cluster size is large enough.

  These results show that the formation of the local magnetic moment in finite
system is different from that in bulk case. Since the size is finite, the
electronic structure and the density of states are modified by the quantum
size effect, the condition of the occurrence for local magnetic
moments in infinite systems, U$\rho_{d\sigma}(\mu) > 1$, might need to be
modified for the formation of local magnetic moment in small system. The condition
should show explicit size dependence, can it be
U(N)$\rho_{d\sigma}(\mu,N) \ > 1$ ? But the finite boundary and the broken
translation-symmetry make it too difficult to obtain an explicit expression in
the momentum space as in the case of infinite system. According to the
experimental results [ 1, 2 ], it seems
that the deviation of the local moment in small systems to that in bulks,
$\delta \mu$, may be the order of $\frac{1}{N}$ in magnitude, i.e.,
 $\delta \mu$ $\sim$ $O(\frac{1}{N})$, when N exceeds a certain critical
value. In our group, some work are being done to try to elucidate the problem by
the first-principle electronic structure calculation for the clusters with the 3d
or 4d transition metal impurity.

The peak and valley of the local magnetic moments
in certain sizes of 1D systems might be related to the magic number of the
total electrons. It should be pointed out that in 1D systems, the scattering
of the conduction electrons by the impurity are dominant, however, near the
absolute zero temperature,
the quantum interference effect will play an important role in 3D case, the
movement of the conduction electrons may be ballistic, and this might
contribute another correction to the formation of the local magnetic moment.

   As a result of the mean-field approach, there are two disadvantages. First
the dynamic fluactuation effect, such as the local spin fluctuation and the
spatial fluctuaction, are neglected; Second, the quantum interference effect
and the weak localization, which may play a role in nanoscale systems, are not
considered. Our next aim is to study the influence of these dynamic and quantum
effects on the formation of local magnetic moments in nanoscale systems.

\begin{center}
{\bf IV. CONCLUSIONS}
\end{center}

    In summary, the magnetic properties of small finite systems up to a few
nanometers have been studied in framework of the Anderson impurity model, we
find that the formation of the local magnetic moment depends not only on the
energy level and the Coulomb interaction of the d-electron but also on the
dimensionality and the
size of finite system, the spectral densities of the local electronic states
and the spatial distribution of the magnetic moment for the polarized
conduction electrons exhibit strong size
dependence both for 1D and 3D systems. These indicate that the difference of
the magnetic behaviors between the small systems and the infinite systems
is not only due to the size effect of exchange interaction between atoms but
also due to the size-dependence of the formation of the local magnetic moments.

\vspace{5mm}

     Zou appreciates Yu Lu and the invitation of the ICTP in Trieste
and Zheng thanks the invitation of the CMIT.
This work is supported by the Climbing Program Research-National Fundamental
Project, the Grant LTWZ of the Chinese Academy of Sciences and the Grant
NSFC of China.

\newpage
\center{\bf REFERENCES}
\begin{enumerate}

\item  I.M.L.Billas, J. A. Becker, A. Chatelian and W. A. de Heer, ~~ Phys. Rev.
        Lett. {\bf 71}, 4067 (1993).
\item  A. J. Cox, J. G. Louderback, S. E. Apsel and L. A. Bloomingfield, ~~
       Phys. Rev. Lett. {\bf 71}, 923 (1993); Phys. Rev. {\bf B49}, 12295 (1994).
\item  D. C. Douglass, A. J. Cox, J. P. Bucher, and L. A. Bloomfield,~~
       Phys. Rev. {\bf B47}, 12874 (1993).
\item  S. N. Khanna and S. Linderoth, ~~ Phys. Rev. Lett. {\bf
       67}, 742 (1991);  J. Magn. Magn. Mater. {\bf 104-107}, 1574 (1992).
\item  G. M. Patsor, R. Hirsch and B. Muhlschlegel,~~ Phys. Rev. Lett. {\bf 72},
       3879 (1994);~~ G. M. Pastor, J. Dorantes-Davila and K.H.Bennemann,~~ Phys.
       Rev. {\bf B40}, 7642 (1989)
\item  B. V. Reddy, S. N.Khanna  and B. I. Dunlap,~~ Phys. Rev. Lett. {\bf 70},
       3323 (1993).
\item  F. Liu, M. R. Press, S. N. Khana, and P. Jena,~~ Phys. Rev. {\bf
       B39}, 6914 (1989); ~~ J. Merikoski, J. Timonen, M. Mannien and P. Jena
       Phys. Rev. Lett. {\bf 66}, 938 (1991).
\item  P. V. Hendriksen, S. Linderoth, and P.-A. Lindgard ,~~ Phys. Rev. {\bf
       B48}, 7259 (1993);  Phys. Rev. {\bf B49}, 12291 (1994).
\item  X. G. Gong and Q. Q. Zheng,~~~J. Phys.: Condens. Matter. {\bf 7},
       2421 (1995).
\item  D. Guenzburger, and D. E. Ellis, ~~~ Phys. Rev. Lett.{\bf 67}, 3832
       (1991);  J. Magn. Magn. Mat. {\bf 104-107}, 2009 (1992).
\item  D. Bagayoko, N. Brener, D. Kanhere and J. Callaway,~~~ Phys. Rev. {\bf
       B36}, 9263 (1987).
\item  P. W. Anderson, ~~ Phys. Rev. {\bf 124}, 41 (1961);~~ Rev. Mod. Phys.
       {\bf B50}, 191 (1978), and some references therein.
\item  J. Friedel,~~ Can. J. Phys. {\bf 34}, 1190 (1958). Nuovo Cimento Suppl.
       VII, 287.  \\
       A. Blandin and J. Friedel,~~J. Phys. Radium  {\bf 20}, 160 (1959)
\item  A. M. Clogston, B. T. Matthias, M. Peter, H. J. Williams, F.
       Corenzwit and R. C. Sherwood, ~~ Phys. Rev. {\bf 125}, 541 (1962).
\item  P. A. Wolff, ~~ Phys. Rev. {\bf 124}, 1030 (1961).
\item  R. M. White, ~~{\it Quantum theory of Magnetism}, ~ Chapt.7, (Springer
       series in Solid State Physics, Vol{\bf 32}), MacGraw-Hill Inc. New York,
       1970.
\item  A. Blandin, {\it Magnetism, ~~Vol.5, ~~ Magnetic Properties of Metallic
       Alloys}, Ed. by H.Suhl, Academic Press, Inc. NY, 1973.
\item  W. J. C. Menezes and M. B. Knickelbein,~~~ Z. Phys.{\bf D26}, 322
       (1993).
\item  J. Kondo, ~~~Prog. Theoret. Phys. Jap. {\bf 32}, 37 (1964).
\item  K. G. Wilson,~~~ Rev. Mod. Phys. {\bf 47}, 773 (1975).
\item  I. K. Yanson, V. V. Fisun, R. Hesper, A.V. Khotkevich, J. M. Krans, J.
       A. Mydosh And J. M. Van Ruitenbeck, ~~~ Phys. Rev. Lett. {\bf 74}, 302
       (1995).

\end{enumerate}

\newpage
\center{\bf FIGURE CAPTIONS}
\begin{enumerate}
\noindent {\bf Fig.1.}  The size-dependence of the local magnetic moments
for several linear quantum chains. Theoretical parameters:
(a) $U$ =8.5, $V$ =0.8, $\epsilon_{d}$ =-2.5,~~~~
(b) $U$ =4.0, $V$ =0.4, $\epsilon_{d}$ =-0.5,~~~~
(c) $U$ =4.0, $V$ =0.8, $\epsilon_{d}$ =-0.5.
\vspace{1cm}

\noindent {\bf Fig.2.}  The dependence of the local magnetic moment on
the size of 1D systems. Parameters:
(a) $U$=4.0, $V$ = 0.4, $\epsilon_{d}$= -1.0, ~~~~
(b) $U$=8.0, $V$ = 0.8, $\epsilon_{d}$= -3.0 .
\vspace{1cm}

\noindent {\bf Fig.3.}  The total spectral densities of the
local electrons for the 1D system with N=15. The parameters are
$U$=8.5, $V$=0.8 and $\epsilon_{d}$=-2.5.
\vspace{1cm}

\noindent {\bf Fig.4.}  The total spectral densities of the
local electrons for the 1D system with N=25. The parameters are
$U$=8.5, $V$=0.8 and $\epsilon_{d}$=-2.5.
\vspace{1cm}

\noindent {\bf Fig.5.}  The spatial distribution of the spins for the
conduction electrons polarized by local moments in the 1D systems
with N=15 (dash line) and with N=25 (solid line). Theoretical parameters
in both two systems: $U$=8.5,
$V$=0.8 and $\epsilon_{d}$=-2.5. The local magnetic moments of the impurity are
$m_{d}=0.945 $$\mu_{B}$ and m$_{d}$=0.976, respectively.
\vspace{1cm}

\noindent {\bf Fig.6.}  The dependence of the local magnetic moment on
the size for the 3D systems. Parameters:
(a) $U$=20.0, $V$ = 0.4, $\epsilon_{d}$= -10.0,
(b) $U$= 3.0, $V$ = 0.8, $\epsilon_{d}$= - 1.5.

\end{enumerate}
\end{document}